\documentclass[12pt,american]{article}
\usepackage[utf8]{inputenc}
\usepackage[a4paper]{geometry}
\usepackage{babel}
\usepackage{amsmath}
\usepackage{amssymb}
\usepackage{graphicx}
\usepackage{setspace}
\usepackage{esint}
\usepackage[authoryear]{natbib}
\PassOptionsToPackage{normalem}{ulem}
\usepackage{ulem}
\usepackage[unicode=true,pdfusetitle,
 bookmarks=true,bookmarksnumbered=false,bookmarksopen=false,
 breaklinks=false,pdfborder={0 0 1},backref=false,colorlinks=false]
 {hyperref}

\makeatletter

\providecommand{\tabularnewline}{\\}

\@ifundefined{date}{}{\date{}}

\usepackage{indentfirst}
\usepackage{babel}
\usepackage{color}
\@ifundefined{definecolor}{color}{}

\usepackage[active]{srcltx}
\usepackage{graphicx}
\usepackage{amsmath}
\makeatletter
\newcommand{\distas}[1]{\mathbin{\overset{#1}{\kern\z@\sim}}}%
\newsavebox{\mybox}\newsavebox{\mysim}
\newcommand{\distras}[1]{%
  \savebox{\mybox}{\hbox{\kern3pt$\scriptstyle#1$\kern3pt}}%
  \savebox{\mysim}{\hbox{$\sim$}}%
  \mathbin{\overset{#1}{\kern\z@\resizebox{\wd\mybox}{\ht\mysim}{$\sim$}}}%
}

\makeatother

\begin{document}

\title{Distance for Functional Data Clustering Based on Smoothing Parameter
Commutation }

\author{ShengLi Tzeng\\
 Department of Public Health\\
 China Medical University\\
 Taiwan\\
 \emph{slt.cmu@gmail.com} \and Christian Hennig\\
Department of Statistical Science\\
University College London\\
United Kingdom\\
\emph{c.hennig@ucl.ac.uk} \and Yu-Fen Li\\
Graduate Institute of Biostatistics\\
China Medical University \\
Taiwan\\
\emph{yufenli@mail.cmu.edu.tw}\and Chien-Ju Lin\\
MRC Biostatistics Unit\\
United Kingdom\\
\emph{chienju@mrc-bsu.cam.ac.uk}}
\maketitle
\begin{abstract}
We propose a novel method to determine the dissimilarity between subjects
for functional data clustering. Spline smoothing or interpolation
is common to deal with data of such type. Instead of estimating the
best-representing curve for each subject as fixed during clustering,
we measure the dissimilarity between subjects based on varying curve
estimates with commutation of smoothing parameters pair-by-pair (of
subjects). The intuitions are that smoothing parameters of smoothing
splines reflect inverse signal-to-noise ratios and that applying an
identical smoothing parameter the smoothed curves for two similar
subjects are expected to be close. The effectiveness of our proposal
is shown through simulations comparing to other dissimilarity measures.
It also has several pragmatic advantages. First, missing values or
irregular time points can be handled directly, thanks to the nature
of smoothing splines. Second, conventional clustering method based
on dissimilarity can be employed straightforward, and the dissimilarity
also serves as a useful tool for outlier detection. Third, the implementation
is almost handy since subroutines for smoothing splines and numerical
integration are widely available. Fourth, the computational complexity
does not increase and is parallel with that in calculating Euclidean
distance between curves estimated by smoothing splines.

\end{abstract}

\section{Introduction}

Clustering sets out to find groups for subjects based on several different
characteristics (variables) with no subgroup labels other than the
observed information. Ideal clustering memberships achieve the target
such that subjects within a cluster are considered to be similar for
the given characteristics (variables). The degree of similarity and
dissimilarity can be defined in plenty of ways, and there are various
methods for grouping subjects, including hierarchical clustering ,
k-means, and DBSCAN to name a few. See e.g. \citet{berkhin2006survey},
\citet{bouveyron2014model}, \citet{murtagh2012algorithms} for brief
literature review of conventional clustering analysis in a multivariate
data context.

In many situations, however, only one variable per subject was measured,
but it was measured time after time. Functional data clustering is
a somewhat distinctive notion to deal with grouping based on such
data. The functional data clustering differs from conventional clustering
in two aspects: data format and time coordinates. First, the data
may be collected at unequally spaced time points, and many ‘missing’
values occur if an analyst aligns records into the conventional ‘variable-by-variable’
format. Second, even all subjects were observed at the same time points,
the conventional clustering fails to take into account the coordinating
order of variables, on which adjacent data collected for the same
subject are expected to have similar values. Several methods for functional
data have been suggested in the literature, and we review three major
categories in the following: distance-based methods, decomposition-based
methods, and model-based methods. 

Distance-based methods, using pointwise distance between pairs of
subjects, are the most straightforward approach (e.g., \citealp{tarpey2003clustering};
\citealp{genolini2010kml}). They often deal with the two issues mentioned
above by certain curve smoothing or imputation techniques, and subsequently
distances between subjects are computed to which the conventional
distance-based methods can be applied.  Little attention, however,
has been paid to the uncertainty of smoothing or imputation. To the
best of our knowledge, the only two exceptions are (1) the prediction
based approach of \citet{alonso2006time} that modified by \citet{vilar2010non},
and (2) the hypotheses-testing-like approach of \citet{maharaj1996significance}.
The former is computationally intensive and the latter is designed
for invertible ARMA process, which restrict their application. 

Decomposition-based methods, overcome the smoothing and sequential
order issues through transforming the observed data into a finite
series of common features, and the procedures deal with uncertainty
of smoothing implicitly. For example, \citet{abraham2003unsupervised}
used spline basis functions, \citet{james2000principal} used functional
principal component analysis, and \citet{warren2005clustering} reviewed
more sophisticated ‘feature-extraction’ algorithms. These approaches
define common features for all groups and then assign weights to features
by which groups are identified. Each group has different weights on
those features and each group can be interpreted according to its
lower-dimensional projection on features. Features extracted from
a certain transformation of data are also popular, such as  spectral
densities (\citealp{fan2004generalised}), periodogram (\citealp{caiado2006periodogram};
\citealp{de2010classification}), and permutation distribution (\citealp{brandmaier2012permutation}).
Nonetheless, in reality not all groups share the same number of features,
and how to determine an appropriate number of dimensions is not easy. 

In light of the difficulties encountered by the first two methods,
many researchers suggest the third alternative, various model-based
frameworks.They estimate individual underlying curves and cluster
subjects simultaneously, and then statistical inference can be made
based on the working models for clusters, such as measuring the uncertainty
for cluster assignment and ‘within-cluster’ variation. Unfortunately,
these approaches encounter other challenges. Purely parametric functional
forms used in traj \citep{jones2007advances} may not be realistic
and its assumption of subjects sharing the same `underlying' curve
within a group can be too restrictive. Applying semi- or non-parametric
methods has to do some dimension reduction within
each group (e.g., FCM by \citealp{james2003clustering}; funHDDC by
\citealp{bouveyron2011model}; Funclust by \citealp{jacques2013funclust};
and K-centre by \citealp{chiou2007functional}), but this encounters
a similar problem as decomposition-based methods. A pure likelihood-based
framework (without dimension reduction) called longclust is proposed
by \citet{mcnicholas2010model}. This method is limited to short time
series and breaks down easily due to the curse of dimensionality.
Even worse, the notion of distribution for random functions is not
well-defined as curves could have infinite dimensions (see e.g.,
\citealp{delaigle2010defining}). 

 The aforementioned review describes the strengths and weaknesses
of the existing functional data clustering methods. Moreover, it is
worth mentioning that the curve variability is an important issue.
Clustering curves can be a difficult ‘chicken-and-egg’ problem between
(1) how to determine the within-cluster variations before identifying
subgroups, and (2) how to separate subgroups when within-cluster variations
are unknown. This dilemma is related directly to the smoothing uncertainty
problem in distance-based approaches. Decomposition-based and model-based
approaches estimate such variability with necessity, but the estimation
id often distorted when outliers occurs. A two-step strategy exploiting
relative merits of different methods seems reasonable: initially separate
potential outliers based on 'outlier-invariant' pairwise distance,
and then form main clusters with another appropriate clustering method.
For such a strategy, a distance measure concerning the variability
of curve estimation or feature selection is crucial. 

In this article, we develop an easily implementable and practically
advantageous method for distance measure between subjects. Instead
of estimating the best-representing curve for each subject as fixed
during clustering, we propose to measure the dissimilarity between
subjects based on pair-by-pair varying curve estimates for a subject.
By applying the technique of smoothing splines, the curve smoothing
is completely determined by the chosen smoothing parameter. The intuitions
behind our proposal are that smoothing parameters of smoothing splines
reflect inverse signal-to-noise ratios and that the smoothing results
for two similar subjects are expected to be close if an identical
smoothing parameter is applied. Specifically, if the unobserved true
curves of subjects $i$ and $j$ are similar, the estimates for them
should resemble with each other, no matter whether we use a smoothing
parameter primarily for the $i$-th or the $j$-th subject. Our distance
is then calculated through commuting between the smoothing parameters
for a pair. 

The rest of the article is organized as follows. Section 2 describes
the proposed dissimilarity and some of its properties. Its effectiveness
is shown through simulations comparing to other dissimilarity measures
in Section 3. An example of its application to methadone dosages observations
is given in Section 4, where we also identified outliers with a rather
simple method. Finally, Section 5 provides some concluding remarks
and discussion concerning future directions.

\section{The Proposed Distance}

We utilize the smoothing spline as our smoothing method, and so we
briefly introduce the smoothing spline before our proposal. Assume
that the curve of $i$-th subject is observed at distinct finite time
points $\{t_{i,1},\ldots,t_{i,K_{i}}\}$ in an interval $[T_{L},T_{U}]$
with measurement errors according to the model 
\begin{equation}
y_{i,k}=f_{i}(t_{i,k})+\epsilon_{i,k},\: k=1,\ldots,K_{i},\: i=1,\ldots,n,\label{eq:observation equation}
\end{equation}
where $\epsilon_{i,k}\distras{i.i.d.}N(0,\sigma^{2})$. A reasonable
estimation of $f_{i}$ is to minimize $\frac{1}{K_{i}}\sum_{k}(y_{i,k}-f_{i}(t_{i,k}))^{2}$
but control the wiggleness of $f_{i}$ such as $\int_{T_{L}}^{T_{U}}(f_{i}^{''}(t))^{2}dt\leq\rho$
for a positive $\rho$. This estimator is equivalent to a smoothing
spline $\hat{f_{i}}(\cdot;\lambda)$ which minimizes 
\begin{equation}
\frac{1}{K_{i}}(\mathbf{y}_{i}-\mathbf{f}_{i})^{\prime}(\mathbf{y}_{i}-\mathbf{f}_{i})+\lambda\int_{T_{L}}^{T_{U}}(f_{i}^{''}(t))^{2}dt\label{eq:spline-variational form}
\end{equation}
given a smoothing parameter $\lambda$, where $\mathbf{y}_{i}=(y_{i,1},\ldots,y_{i,K_{i}})^{\prime}$
and $\mathbf{f}_{i}=\left(f_{i}(t_{i,1}),\ldots,f_{i}(t_{i,K_{i}})\right)^{\prime}$
(see e.g. \citealp{wahba1980some[TPS]}; \citealp{green1993nonparametric}).
There are various methods to determine an appropriate $\lambda$ in
(\ref{eq:spline-variational form}) , and once $\lambda$ chosen $\hat{f}_{i}(t;\lambda)$
for $t\in[T_{L},T_{U}]$ is completely established. We exploit a mixed-effects
model representation (e.g., \citealp{wang1998smoothing}) of the problem
in (\ref{eq:spline-variational form}) as
\begin{equation}
\mathbf{y}_{i}=\mathbf{X}_{i}\boldsymbol{\beta}_{i}+\mathbf{u}_{i}+\boldsymbol{\epsilon}_{i},\label{eq:spline-mixed form}
\end{equation}
where $\boldsymbol{\beta}_{i}$ is the fixed effect, $\mathbf{X}_{i}$
has two columns being $1$'s and $(t_{i,1},\ldots,t_{i,K_{i}})^{\prime}$,
$\boldsymbol{\epsilon}_{i}=(\epsilon_{i,1},\ldots,\epsilon_{i,K_{i}})^{\prime}\sim N(0,\sigma^{2}\mathbf{I})$,
and $\mathbf{u}_{i}\sim N(\mathbf{0},\sigma_{u}^{2}\mathbf{R})$ with
$\sigma_{u}^{2}=\sigma^{2}/(K_{i}\lambda)$ and the $(k,k^{*})$ element
of $\mathbf{R}$ being 
\[
\left(T_{U}-T_{L}\right)^{-2}\int_{T_{L}}^{T_{U}}(t_{i,k}-\tau)_{+}(t_{i,k^{*}}-\tau)_{+}d\tau
\]
with $a_{+}=\max(0,a)$. As a function of variance for $\mathbf{u}_{i}$
in (\ref{eq:spline-mixed form}), $\lambda$ can be determined based
on the restricted maximum likelihood method and $K_{i}\lambda$ has
a useful interpretation of inverse signal-to-noise ratio as $\sigma_{u}^{2}/\sigma^{2}$.
Additionally, it been shown that the smoothing results are more robust
 even when the correlation structure of $\textrm{var}(\boldsymbol{\epsilon}_{i})$
is mis-specified (e.g.\citealp{wang1998smoothing} and \citealp{krivobokova2007note}).

Our proposal starts with finding $\hat{\lambda}_{i}$ in (\ref{eq:spline-mixed form})
for each subject based on $\mathbf{y}_{i}$. The estimated curve is
denoted by $\hat{f}_{i}(\cdot;\hat{\lambda}_{i})$, which amounts
to obtaining $\hat{f}_{i}(\cdot;\lambda)$ given $\lambda=\hat{\lambda}_{i}$
in (\ref{eq:spline-variational form}) for observations $\mathbf{y}_{i}$.
Fixed on the smoothing parameter $\hat{\lambda}_{i}$, we can obtain
$\hat{f}_{j}(\cdot;\hat{\lambda}_{i})$ based on observations $\mathbf{y}_{j}$.
The roles of the two subjects can be exchanged, and similarly we have
$\hat{f}_{j}(\cdot;\hat{\lambda}_{j})$ and $\hat{f}_{i}(\cdot;\hat{\lambda}_{j})$
. Then the distance between subjects $i$ and $j$ is calculated as
\begin{equation}
d_{i,j}=\frac{1}{2}\left\{ \left[\int_{T_{L}}^{T_{U}}\left(\hat{f}_{i}(t;\hat{\lambda}_{i})-\hat{f}_{j}(t;\hat{\lambda}_{i})\right)^{2}dt\right]^{1/2}+\left[\int_{T_{L}}^{T_{U}}\left(\hat{f}_{i}(t;\hat{\lambda}_{j})-\hat{f}_{j}(t;\hat{\lambda}_{j})\right)^{2}dt\right]^{1/2}\right\} .\label{eq:proposed distance}
\end{equation}
Due to the roles of $\hat{\lambda}_{i}$ and $\hat{\lambda}_{j}$
in (\ref{eq:proposed distance}), we call it a smoothing parameter
commutation based distance, and explain its underlying rationale below.
First if the `true' $f_{i}$ and $f_{j}$ are similar, it is expected
that $\hat{f}_{i}$ and $\hat{f}_{j}$ from $\mathbf{y}_{i}$ and
$\mathbf{y}_{j}$ should be close, given an identical smoothing parameter.
Second it takes the variation of smoothing into consideration with
diverse $\lambda$'s for different pair of $(i,j)$'s. It focuses
on how similar a pair of curves can be, instead of the distance between
(fixed) estimated curves. Third $d_{i,j}\geq0$, $d_{i,j}=0$ if $i=j$,
and $d_{i,j}=d_{j,i}$, so conventional distance based clustering
methods can be applied straightforward. Fourth it reduces to rooted
integral squared difference of $f_{i}$ and $f_{j}$ when no missing
values and measurement errors present. 

Our proposal also has several pragmatic advantages. First, missing
values or irregular time points can be handled directly, thanks to
the nature of smoothing splines. Second, the dissimilarity also serves
as a useful tool for outlier detection (see Section \ref{sec:Real-Data-Application}).
Third, the implementation is almost handy since subroutines for smoothing
splines and numerical integration are widely available. Although the
computing burden for (\ref{eq:proposed distance}) seems heavy at
first glance, it can be done more efficiently among $n$ subjects.
Given $\lambda$ a fast $O(K_{i})$ algorithm to compute $\hat{f_{i}}(t;\lambda)$
does exist (e.g., \citealp{hutchinson1985smoothing}). Thus, one needs
to solve $\hat{\lambda}_{i}$ in (\ref{eq:spline-mixed form}) only
$n$ times for the $n$ subjects, and then adopts the fast algorithm
for $\left\{ \hat{f}_{j}(t;\hat{\lambda}_{i}):\: i,j=1,\ldots,n\right\} $.
Therefore the computational complexity is proportional to that in
treating $f=\hat{f_{i}}(t;\lambda_{i})$ as fixed and calculating
distance as squared root of $\int_{T_{L}}^{T_{U}}\left(\hat{f}_{i}(t;\hat{\lambda}_{i})-\hat{f}_{j}(t;\hat{\lambda}_{j})\right)^{2}dt$
(see \citealp{ramsay2005smoothing} and the latter procedure is referred
to as $d_{SS}$ in what follows).

\section{Simulation}

We conduct a simulation to investigate whether our proposed measure
is more representative than other dissimilarity measures when observations
were contaminated with (independent or dependent) noises. If an analyst is
interested in the relative shape pattern of curves, regardless of
shift, shrinkage, expansion, or magnitude, then several alignment,
normalization, and warping tools can be applied in preprocessing (e.g.,\citet{berndt1994using},
\citet{gaffney2004joint}, and \citet{liu2009simultaneous}). For
fear of losing focus, we do not consider distance measures engaging
with the preprocessing. 

We consider the following four random curve models over $t\in[0,1]$
\begin{align*}
f^{(1)}(t;\eta)= & \eta,\\
f^{(2)}(t;\eta)= & \sin(2\pi t)-t+2\eta\cos(4\pi t),\\
{\normalcolor f^{(3)}}(t;\eta)= & 3t+2\eta t,\\
f^{(4)}(t;\eta)= & 5\eta\left\{ (t-0.5)^{2}-2t(1-t)\right\} ,
\end{align*}
where $\eta\sim N(1,0.3^{2})$. The four functional forms stand for
constant, periodic, linear, and nonlinear (unobserved) true curves,
respectively. The observed data are generated according to (\ref{eq:observation equation})
merely at 200 time points, $t_{k}\in\{0,1/199,\ldots,198/199,1\}$,
with noises coming from four mechanisms
\begin{align}
\textrm{WN:\qquad\quad\quad} & \epsilon_{k}=\xi_{k},\nonumber \\
\textrm{AR}:\qquad\quad\quad & \epsilon_{k}=0.8\epsilon_{k-1}+\xi_{k},\nonumber \\
\textrm{SARMA:\quad}\:\:\:\: & \epsilon_{k}=0.8\epsilon_{k-10}+0.8\xi_{k-10}+\xi_{k},\nonumber \\
\textrm{BILR{\normalcolor :}}\qquad\quad\: & \epsilon_{k}=0.8\epsilon_{k-1}+0.2\xi_{k-1}-0.2\epsilon_{k-1}\xi_{k-1}+\xi_{k},\label{eq:noise}
\end{align}
where $\xi_{k}\distras{i.i.d.}N(0,1)$ and $\xi_{k}$ is independent
of $\epsilon_{k^{\prime}}$ for $k^{\prime}\neq k$. That is, we set
$K_{i}\equiv200$, $t_{ik}\equiv(k-1)/199$. The four noise mechanisms
are  examples of usual assumption for noises: purely
independent process, stationary process, cyclostationary process,
and nonstationary process. For each combination of $f\in\left\{ f^{(1)},f^{(2)},f^{(3)},f^{(4)}\right\} $
and mechanism of $\epsilon_{k}$, 10 series are generated according
to 10 independent $\eta$ as well as 10 sets of $\epsilon_{k}$'s,
and totally there are 160 series mimicking the longitudinal observations
from 160 subjects. 

Then several distance measures are calculated based on the simulated
data. Following the notation in \citet{montero2014tsclust}, we compare
10 measures, including our proposal (referred to as $d_{OUR}$) and
point-wise Euclidean distance $d_{EUCL}=\sqrt{\sum_{k}(y_{ik}-y_{jk})^{2}}$,
and the eight others are listed in Table \ref{tab:Distance-measures}.
Two comparison criteria are defined as follows: 
\begin{align*}
Q & =\underset{a,b}{\min}\sum_{i}\sum_{j\neq i}\frac{\left(a+b\hat{d}_{i,j}-d_{i,j}\right)^{2}}{d_{i,j}},\\
R & =\sum_{i}\sum_{j\neq i}(\hat{r}_{i,j}-r_{i,j})^{2},
\end{align*}
where $\hat{d}_{i,j}$ is one of the considered distance measures
between the $i$-th and $j$-th subjects, $d_{i,j}=\sqrt{\sum_{k}(f_{i}(t_{ik})-f_{j}(t_{jk}))^{2}}$
is the true distance without noise, and $\hat{r}_{i,j}$ and $r_{i,j}$
are the corresponding rank of $\hat{d}_{i,j}$ and $d_{i,j}$ among
all pairs of $(i,j)$'s, respectively. The quantity $Q$ reflects
the loss, normalized by the true distance scales, for (linear) approximation
to all the pairs of true distances, while $R$ measures the deviation
from monotonicity between $\hat{d}_{i,j}$ and $d_{i,j}$. A good
measure should have a small value of $Q$ or $R$. The averaged $Q$
and $R$ values for the 10 measures over 200 simulation replicates
are given in Table \ref{tab:Q comparison} and Table \ref{tab:R comparison},
respectively.

The two comparison criteria are highly coherent in that they almost
always sort the same best and worst measures. As expected, $d_{EUCL}$
is often among the best measures since there are no missing data and
$d_{EUCL}$ is unbiased in many situations. But it does not good enough
if the signal or noise is periodic ($f^{(2)}$, SARMA, respectively).
Our method and $d_{SS}$ always fall in the best 3 measures, either
for 10 curves within an individual group or for 160 curves as a whole.
Note that $d_{SS}$ and $d_{EUCL}$ have almost identical result within
a group, due to both utilize the mixed-effects model representation
of smoothing splines. The difference lies in that $d_{SS}$ regarding
$\hat{f}_{i}(t;\hat{\lambda}_{i})$ as a fixed estimate of $f_{i}$.
Our method outperforms for between-group distance, which indicates
the advantage of accounting for smoothing variation via smoothing
parameter commutation. In certain cases $d_{PRED,h}$ and $d_{MAH}$
are good measures, which also take estimation uncertainty into consideration.

\begin{table}
\small	
\begin{centering}
\begin{tabular}{lll}
Notation & Description & Literature\tabularnewline
\hline 
$d_{MAH}$ & parametric testing of equality of processes & \citet{maharaj1996significance}\tabularnewline
$d_{GLK}$ & nonparametric equality testing of log-spectra & \citet{fan2004generalised}\tabularnewline
$d_{SS}$ & based on spline smoothing curves & \citet{ramsay2005smoothing}\tabularnewline
$d_{CORT}$ & correlation-based modification of $d_{EUCL}$ & \citet{chouakria2007adaptive}\tabularnewline
$d_{IP}$ & based on integrated periodogram & \citet{de2010classification}\tabularnewline
$d_{PRED,h}$ & based on predicted values at future & \citet{vilar2010non}\tabularnewline
$d_{CID}$ & complexity-based modification of $d_{EUCL}$ & \citet{batista2011complexity}\tabularnewline
$d_{PDC}$ & permutation distributions of order patterns & \citet{brandmaier2012permutation}\tabularnewline
\hline 
\end{tabular}
\par\end{centering}

\caption{\label{tab:Distance-measures}Distance measures to be compared.}
\end{table}

\begin{table}
\small
\begin{tabular}{crrrrrrrrrr}
\hline 
 & $d_{EUCL}$ & $d_{OUR}$ & $d_{MAH}$ & $d_{GLK}$ & $d_{SS}$ & $d_{CORT}$ & $d_{IP}$ & $d_{PRED,h}$ & $d_{CID}$ & $d_{PDC}$\tabularnewline
\hline 
$f^{(1)}$+W & \textbf{1.59} & \textbf{0.36} & 8.45 & 8.55 & \textbf{0.37} & 3.82 & 9.32 & 1.85 & 2.19 & 8.63\tabularnewline
$f^{(1)}$+A & \textbf{5.61} & \textbf{5.66} & 8.07 & 8.02 & \textbf{5.66} & 6.4 & 9.54 & \textbf{4.61} & 5.82 & 7.96\tabularnewline
$f^{(1)}$+S & 8.07 & \textbf{6.15} & 8.67 & 8.55 & \textbf{6.16} & 8.68 & 10.53 & \textbf{7.32} & 8.19 & 8.78\tabularnewline
$f^{(1)}$+B & \textbf{7.65} & \textbf{7.66} & 8.48 & 8.48 & \textbf{7.65} & 7.89 & 11.88 & \textbf{5.62} & 7.87 & 8.48\tabularnewline
$f^{(2)}$+W & 2.21 & \textbf{0.87} & 3.79 & 3.56 & \textbf{0.83} & 3.54 & \textbf{1.35} & 3.96 & 2.76 & 4.06\tabularnewline
$f^{(2)}$+A & \textbf{3.94} & \textbf{3.94} & \textbf{3.91} & 3.97 & \textbf{3.94} & \textbf{3.94} & 5.69 & 4.01 & \textbf{3.95} & 3.97\tabularnewline
$f^{(2)}$+S & 3.96 & \textbf{3.81} & 3.83 & \textbf{3.83} & \textbf{3.63} & 3.94 & 5.71 & 4.04 & 3.93 & 3.95\tabularnewline
$f^{(2)}$+B & \textbf{3.99} & \textbf{3.99} & 4.05 & 4.06 & \textbf{3.99} & \textbf{3.99} & 12.32 & 4.04 & 4.04 & 4.09\tabularnewline
$f^{(3)}$+W & 1.49 & \textbf{1.05} & 1.40 & 1.40 & \textbf{0.99} & 1.52 & \textbf{1.38} & 1.58 & 1.50 & 1.53\tabularnewline
$f^{(3)}$+A & \textbf{1.49} & \textbf{1.49} & \textbf{1.47} & \textbf{1.49} & \textbf{1.49} & 1.50 & 2.38 & 1.51 & \textbf{1.49} & \textbf{1.49}\tabularnewline
$f^{(3)}$+S & 1.53 & \textbf{1.49} & \textbf{1.49} & \textbf{1.49} & \textbf{1.49} & 1.52 & 4.07 & 1.54 & 1.52 & 1.51\tabularnewline
$f^{(3)}$+B & \textbf{1.56} & \textbf{1.56} & \textbf{1.55} & \textbf{1.56} & \textbf{1.56} & 1.57 & 12.44 & 1.58 & \textbf{1.56} & \textbf{1.56}\tabularnewline
$f^{(4)}$+W & \textbf{2.31} & \textbf{0.79} & 3.17 & 3.19 & \textbf{0.81} & 2.94 & 3.53 & 2.69 & 2.54 & 3.18\tabularnewline
$f^{(4)}$+A & \textbf{3.20} & \textbf{3.20} & 3.21 & 3.27 & \textbf{3.20} & 3.23 & 4.01 & \textbf{3.04} & 3.23 & 3.22\tabularnewline
$f^{(4)}$+S & 3.29 & \textbf{3.19} & 3.23 & \textbf{3.22} & \textbf{3.18} & 3.29 & 4.74 & 3.32 & 3.28 & 3.25\tabularnewline
$f^{(4)}$+B & \textbf{3.35} & \textbf{3.35} & 3.38 & 3.38 & \textbf{3.35} & 3.36 & 9.01 & 3.31 & 3.38 & 3.37\tabularnewline
ALL & \textbf{24.83} & \textbf{23.92} & 29.29 & 29.30 & \textbf{24.45} & 26.13 & 32.38 & 25.63 & 28.86 & 29.28\tabularnewline
\hline 
\end{tabular}\caption{\label{tab:Q comparison} Averaged Q values over 200 simulated replicates
among 10 distance measures for each combination of $f$ and $\epsilon_{k}$
(with 10 random curves), and all the 160 curves. W, A, S, and B in
the first column stand for WN, AR, SARMA, and BILR in (\ref{eq:noise}),
respectively. Bold digits are the best 3 within each row.}
\end{table}

\begin{table}
\footnotesize
\begin{tabular}{crrrrrrrrrr}	
\hline 
 & $d_{EUCL}$ & $d_{OUR}$ & $d_{MAH}$ & $d_{GLK}$ & $d_{SS}$ & $d_{CORT}$ & $d_{IP}$ & $d_{PRED,h}$ & $d_{CID}$ & $d_{PDC}$\tabularnewline
\hline 
$f^{(1)}$+W & \textbf{0.73} & \textbf{0.24} & 12.26 & 12.15 & \textbf{0.24} & 2.22 & 12.01 & 1.11 & 1.16 & 12.39\tabularnewline
$f^{(1)}$+A & \textbf{4.89} & \textbf{4.89} & 12.11 & 12.02 & \textbf{4.89} & 5.85 & 12.29 & \textbf{3.98} & 5.15 & 12.29\tabularnewline
$f^{(1)}$+S & 7.79 & \textbf{5.21} & 11.82 & 11.94 & \textbf{5.24} & 10.20 & 12.29 & \textbf{7.23} & 8.87 & 12.18\tabularnewline
$f^{(1)}$+B & \textbf{7.73} & \textbf{7.73} & 12.27 & 12.15 & \textbf{7.73} & 8.30 & 12.25 & \textbf{4.70} & 8.20 & 12.43\tabularnewline
$f^{(2)}$+W & 3.01 & \textbf{1.04} & 8.88 & 6.69 & \textbf{1.01} & 6.27 & \textbf{1.29} & 11.69 & 4.21 & 12.27\tabularnewline
$f^{(2)}$+A & \textbf{9.20} & \textbf{9.20} & 10.24 & 10.59 & \textbf{9.19} & 9.80 & 8.15 & 12.66 & 9.45 & 12.05\tabularnewline
$f^{(2)}$+S & 10.99 & \textbf{8.19} & 10.14 & 10.18 & \textbf{7.88} & 11.71 & \textbf{7.85} & 13.45 & 11.33 & 12.35\tabularnewline
$f^{(2)}$+B & \textbf{10.59} & \textbf{10.6} & 11.62 & 11.63 & \textbf{10.6} & 10.89 & 10.77 & 12.66 & 10.75 & 12.10\tabularnewline
$f^{(3)}$+W & 9.18 & \textbf{4.49} & 8.16 & 8.10 & \textbf{4.27} & 10.79 & \textbf{6.78} & 14.54 & 10.09 & 12.35\tabularnewline
$f^{(3)}$+A & \textbf{11.5} & \textbf{11.53} & 11.89 & 11.87 & \textbf{11.53} & 11.69 & 12.06 & 13.88 & \textbf{11.53} & 12.22\tabularnewline
$f^{(3)}$+S & 11.90 & \textbf{11.53} & \textbf{11.72} & 12.17 & \textbf{11.34} & 11.95 & 12.10 & 14.09 & 11.87 & 11.96\tabularnewline
$f^{(3)}$+B & \textbf{11.99} & \textbf{12.02} & 12.12 & 12.05 & \textbf{12.02} & \textbf{11.94} & 12.26 & 13.51 & 12.06 & 12.31\tabularnewline
$f^{(4)}$+W & \textbf{4.63} & \textbf{1.31} & 11.56 & 12.23 & \textbf{1.32} & 7.87 & 12.29 & 7.49 & 5.79 & 12.21\tabularnewline
$f^{(4)}$+A & \textbf{9.89} & \textbf{9.89} & 11.59 & 12.28 & \textbf{9.88} & 10.45 & 12.47 & 10.00 & 10.02 & 12.18\tabularnewline
$f^{(4)}$+S & 11.71 & \textbf{10.23} & \textbf{11.34} & 12.11 & \textbf{10.23} & 11.87 & 12.19 & 13.08 & 11.72 & 12.20\tabularnewline
$f^{(4)}$+B & \textbf{11.24} & \textbf{11.26} & 12.24 & 12.20 & \textbf{11.26} & 11.52 & 12.24 & \textbf{10.83} & 11.41 & 11.68\tabularnewline
ALL & \textbf{1155.6}  & \textbf{874.7}  & 4160.5  & 4063.7  & \textbf{901.0 } & 1315.4  & 3768.6  & 1239.8  & 2693.1  & 4191.6 \tabularnewline
\hline 

\end{tabular}\caption{\label{tab:R comparison}Averaged R values over 200 simulated replicates
among 10 distance measures for each combination of $f$ and $\epsilon_{k}$
(with 10 random curves), and all the 160 curves. W, A, S, and B in
the first column stand for WN, AR, SARMA, and BILR in (\ref{eq:noise}),
respectively. Bold digits are the best 3 within each row.}
\end{table}

\section{Real Data Application\label{sec:Real-Data-Application}}

We shall apply (\ref{eq:proposed distance}) to a methadone maintenance
therapy data in \citet{lin2015clustering}. Daily methadone dosages
in mg for 314 participants between 01 January 2007 and 31 December
2008 were collected. The (partially) observed dose levels for each
patient from day 1 to day 180 were used for clustering. \citet{lin2015clustering}
categorized the dosages into 7 levels, one of which is missing value,
and proposed a new dissimilarity measure for clustering ordinal data.
The ordering of time coordinates, however, were discarded in their
approach. In this example, we use the primary prescription dosage,
and do not recode missing values separately. Smoothing splines take
care the irregular follow-up time points of patients automatically,
which may not be an easy task for other measures listed in Table \ref{tab:Distance-measures}.

\begin{figure}
\begin{centering}
\includegraphics[scale=0.5]{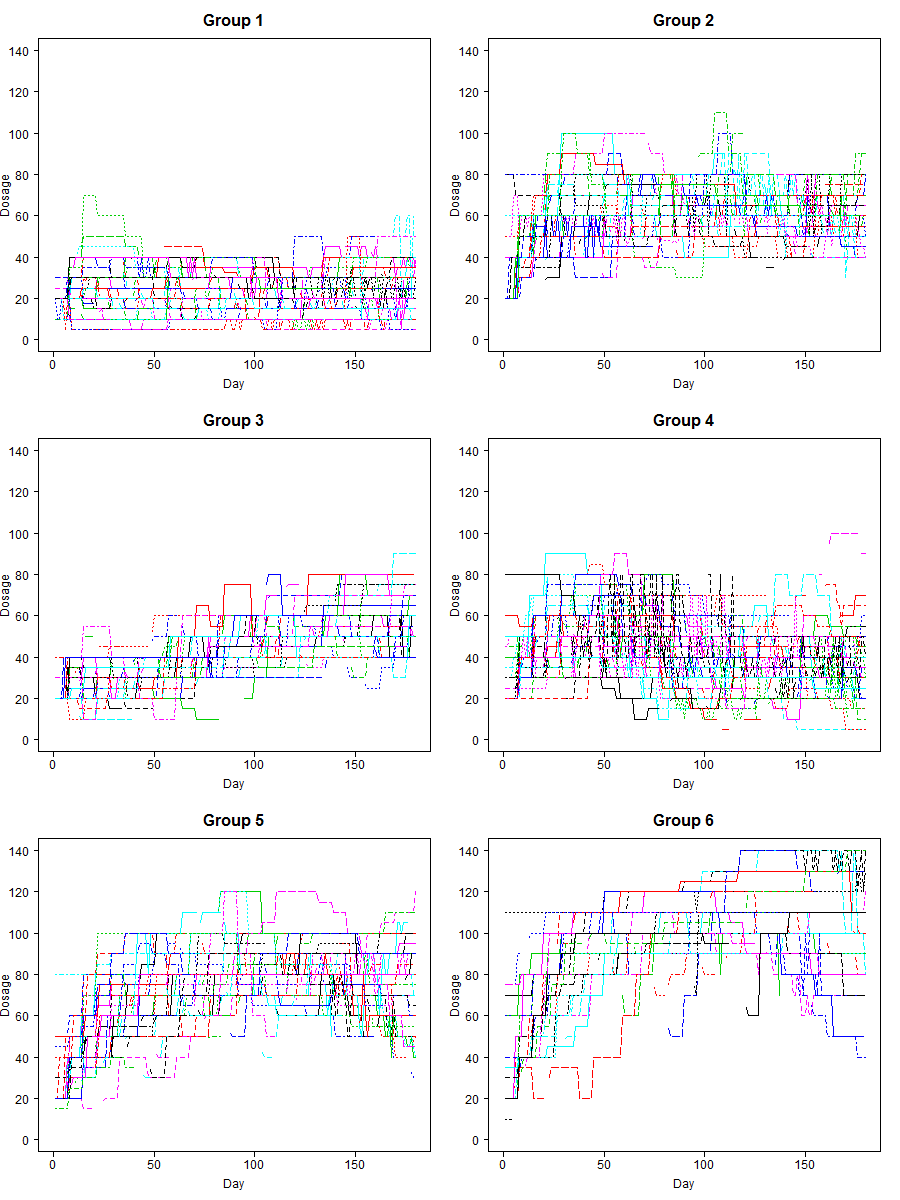}
\par\end{centering}

\caption{\label{fig:Subgroups}Subgroups from PAM clustering of the 312 patients
in methadone maintenance therapy.}
\end{figure}

\begin{figure}
\noindent \begin{centering}
\centering{}%
\begin{tabular}{cc}
\includegraphics[scale=0.48]{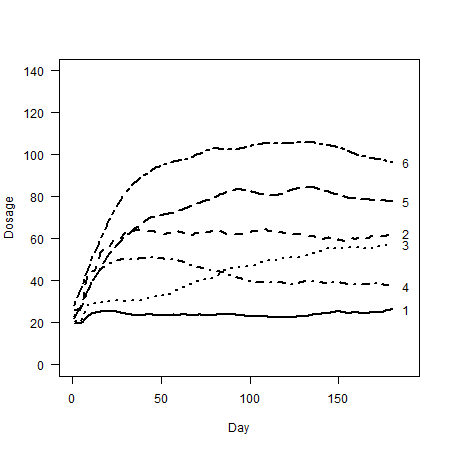}  & \includegraphics[scale=0.48]{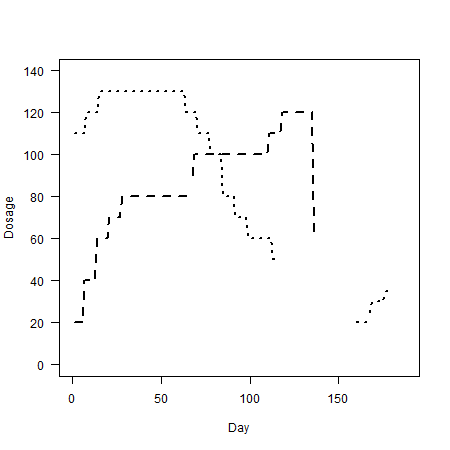}\tabularnewline
(a)  & (b)\tabularnewline
\includegraphics[scale=0.48]{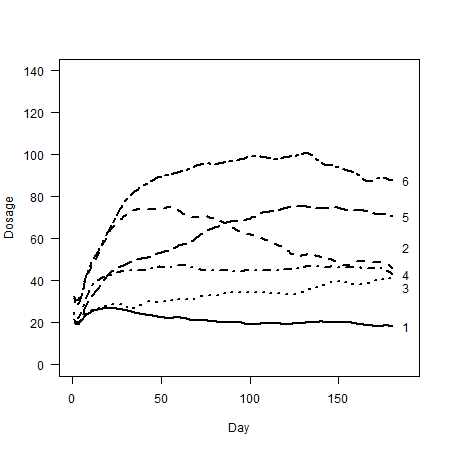} & \includegraphics[scale=0.48]{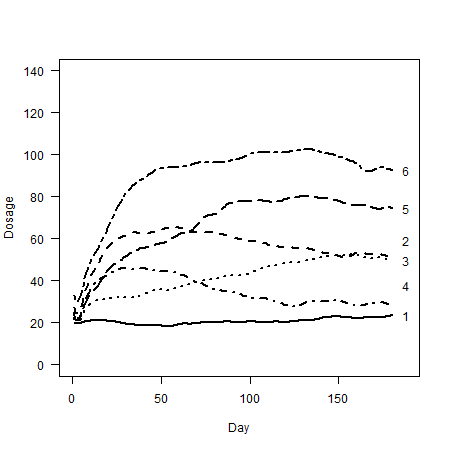}\tabularnewline
(c)  & (d)\tabularnewline
\end{tabular}
\par\end{centering}

\noindent \centering{}\medskip{}
 \caption{\label{fig:mean-profile}(a): Mean curves of subgroups in Figure \ref{fig:Subgroups};
(b) Dosage profiles of the two excluded outliers; (c) Mean profiles
of a model-based clustering method including the two outliers; (d)
Mean profiles of the same clustering method with (c) but excluding
the two outliers. }

\label{fig:prediction map} 
\end{figure}

The clustering procedure consists of three steps: (1) calculating
the distance matrix , (2) detecting and removing outliers, and (3)
forming clusters with the remaining data. We started from obtaining
the pairwise distance matrix based on (\ref{eq:proposed distance}).
Then two outliers were simply detected by calculating the average
distance of each patient's nearest 3 neighbors. Two had the distance
in magnitude of 500 and 1 010, while all the others had distance falling
{[}39,300{]}. Cluster identification result can be affected significantly
by a few far away noisy points, which should be removed in order to
make more reliable clustering. Our method to detect outliers is similar
to \citet{ramaswamy2000efficient} based on dissimilarity. Excluding
the two outliers, the remaining 312 dosage curves of patients were
clusterd into 5 subgroups via “partitioning around medoids” (PAM),
as shown in Figure \ref{fig:Subgroups}. The mean curves for each
subgroups are also shown in Figure \ref{fig:mean-profile} (a). 

It is obvious Group 1 and 2 are more stable, remaining a dose level
roughly within {[}10,40{]} and {[}40,80{]}, respectively. Group 3
has an upward trend while Group 4 has a downward trend, and from Figure
\ref{fig:mean-profile} the two mean curves cross around day 85. Group
5 goes up quickly and stay a dose level around 80. Although Group
6 has a similar trend to Group 5, it fluctuates heavily over a larger
range and looks more unstable. Overall, these figures indicate that
a patient with early higher dosage taken (roughly above 60 mg at day
45) tends not to reduce the level afterward and a monitoring between
the second and third month can be critical. 

Results based on a model-based functional data clustering are also
given for comparison. We used the `funcit' function in the 'funcy'
package (\citealp{funcy}) on The Comprehensive R Archive Network
(CRAN; \citealp{CRAN}). The model option of the function is set to
be `iterSubspace', i.e., an implementation of the algorithm in \citealp{chiou2007functional}.
The theoretical mean profiles of clusters based on participants including
and excluding outliers are shown in Figure \ref{fig:mean-profile}
(c) and Figure \ref{fig:mean-profile} (d), respectively. Profiles
of the two outlier participants are also shown in Figure \ref{fig:mean-profile}
(b). 

Although PAM does not provide theoretical mean profiles so that it
can not be directly compared to the model-based method, note the resemblance
between Figures \ref{fig:mean-profile} (a) and \ref{fig:mean-profile}
(d). Excluding the two outliers did improve the model-based method
in that the average distance to mean profile reduced 7.6\% from 166.7
to 154.9, which gave more compact clusters. Inspecting Figure \ref{fig:mean-profile}
(b), we can realize the interlacing of the 2nd, 3rd, and 4th subgroups
in Figure \ref{fig:mean-profile} (c). Clearly, it is hard to group
the two curves of outliers into the found groups. Forcing to include
them needs to exaggerate the within-group variation, no matter which
groups they are assigned to. Then the boundaries of groups are getting
blurred, so are the representativeness of mean profiles. 

Unfortunately, identifying outlier during the model-based clustering
procedure can be tautological, since the unknown `ordinary' within-group
variation depends on telling apart which are `abnormal' participants.
In contrast, dissimilarity in a distance-based method (including our
proposal) is not affected by whether outliers occurs, and can serve
as an outlier detector. The simulations above reveal the stable superiority
of the proposed dissimilarity, and it is usable in a beneficial preclean
step for model-based clusterings.


\section{Conclusion and Discussion}

We have shown that distance based on smoothed data is better than
distance based on specific time series assumptions, if the underlying
curves are changed gradually. With smoothing parameter commutation,
the proposed distance measure gains some improvement of the widely
used approach in \citet{ramsay2005smoothing} without introducing
further computational complexity. We also demonstrated a simple method
for outlier detection that helps model-based functional data clustering
form more compact subgrous.

The `funcy' package on CRAN integrated several model-based clustering
methods for functional data, but most of them require regular measurements
and do not fit the methadone dosage example with many missing values.
The only two methods of the package allowing irregular measurements
are `fitfclust' and `iterSubspace', and we apply the latter merely
because the former was eating up more than 20GB memories and spending
6 hours at each iteration for the example, which is not yet a practical
choice for general applications. 

There are many other nonparametric regression methods other than smoothing
splines, e.g., local polynomial regressions, wavelet analysis. Different
techniques stand out in different situations. It is of interest to
study whether there exist analogous parameter commutation operations
and similar advantages when applying other nonparametric regressions.
This direction is left as a future work. 


\bibliography{fdclust-arxiv}

\begin{thebibliography}{}

\bibitem[\protect\citeauthoryear{Abraham, Cornillon, Matzner-L{\o}ber, and
  Molinari}{Abraham et~al.}{2003}]{abraham2003unsupervised}
Abraham, C., P.-A. Cornillon, E.~Matzner-L{\o}ber, and N.~Molinari (2003).
\newblock Unsupervised curve clustering using {B}-splines.
\newblock {\em Scandinavian journal of statistics\/}~{\em 30\/}(3), 581--595.

\bibitem[\protect\citeauthoryear{Alonso, Berrendero, Hern{\'a}ndez, and
  Justel}{Alonso et~al.}{2006}]{alonso2006time}
Alonso, A.~M., J.~R. Berrendero, A.~Hern{\'a}ndez, and A.~Justel (2006).
\newblock Time series clustering based on forecast densities.
\newblock {\em Computational Statistics and Data Analysis\/}~{\em 51\/}(2),
  762--776.

\bibitem[\protect\citeauthoryear{Batista, Wang, and Keogh}{Batista
  et~al.}{2011}]{batista2011complexity}
Batista, G.~E., X.~Wang, and E.~J. Keogh (2011).
\newblock A complexity-invariant distance measure for time series.
\newblock In {\em Proceedings of the 11th SIAM International Conference on Data
  Mining}, pp.\  699--710.

\bibitem[\protect\citeauthoryear{Berkhin}{Berkhin}{2006}]{berkhin2006survey}
Berkhin, P. (2006).
\newblock A survey of clustering data mining techniques.
\newblock In J.~Kogan, C.~Nicholas, and M.~Teboulle (Eds.), {\em Grouping
  multidimensional data}, pp.\  25--71. Springer.

\bibitem[\protect\citeauthoryear{Berndt and Clifford}{Berndt and
  Clifford}{1994}]{berndt1994using}
Berndt, D.~J. and J.~Clifford (1994).
\newblock Using dynamic time warping to find patterns in time series.
\newblock In {\em KDD-94: AAAI Workshop on Know ledge Dis- covery in
  Databases}, Volume~10, pp.\  359--370. Seattle, Washington.

\bibitem[\protect\citeauthoryear{Bouveyron and Brunet-Saumard}{Bouveyron and
  Brunet-Saumard}{2014}]{bouveyron2014model}
Bouveyron, C. and C.~Brunet-Saumard (2014).
\newblock Model-based clustering of high-dimensional data: A review.
\newblock {\em Computational Statistics and Data Analysis\/}~{\em 71}, 52--78.

\bibitem[\protect\citeauthoryear{Bouveyron and Jacques}{Bouveyron and
  Jacques}{2011}]{bouveyron2011model}
Bouveyron, C. and J.~Jacques (2011).
\newblock Model-based clustering of time series in group-specific functional
  subspaces.
\newblock {\em Advances in Data Analysis and Classification\/}~{\em 5\/}(4),
  281--300.

\bibitem[\protect\citeauthoryear{Brandmaier}{Brandmaier}{2012}]{brandmaier2012permutation}
Brandmaier, A.~M. (2012).
\newblock {\em Permutation distribution clustering and structural equation
  model trees}.
\newblock Ph.\ D. thesis, Saarland University, Saarbruecken, Germany.

\bibitem[\protect\citeauthoryear{Caiado, Crato, and Pe{\~n}a}{Caiado
  et~al.}{2006}]{caiado2006periodogram}
Caiado, J., N.~Crato, and D.~Pe{\~n}a (2006).
\newblock A periodogram-based metric for time series classification.
\newblock {\em Computational Statistics and Data Analysis\/}~{\em 50\/}(10),
  2668--2684.

\bibitem[\protect\citeauthoryear{Chiou and Li}{Chiou and
  Li}{2007}]{chiou2007functional}
Chiou, J.-M. and P.-L. Li (2007).
\newblock Functional clustering and identifying substructures of longitudinal
  data.
\newblock {\em Journal of the Royal Statistical Society: Series B (Statistical
  Methodology)\/}~{\em 69\/}(4), 679--699.

\bibitem[\protect\citeauthoryear{Chouakria and Nagabhushan}{Chouakria and
  Nagabhushan}{2007}]{chouakria2007adaptive}
Chouakria, A.~D. and P.~N. Nagabhushan (2007).
\newblock Adaptive dissimilarity index for measuring time series proximity.
\newblock {\em Advances in Data Analysis and Classification\/}~{\em 1\/}(1),
  5--21.

\bibitem[\protect\citeauthoryear{De~Lucas}{De~Lucas}{2010}]{de2010classification}
De~Lucas, D.~C. (2010).
\newblock {\em Classification techniques for time series and functional data}.
\newblock Ph.\ D. thesis, Universidad Carlos III de Madrid.

\bibitem[\protect\citeauthoryear{Delaigle and Hall}{Delaigle and
  Hall}{2010}]{delaigle2010defining}
Delaigle, A. and P.~Hall (2010).
\newblock Defining probability density for a distribution of random functions.
\newblock {\em The Annals of Statistics\/}~{\em 38\/}(2), 1171--1193.

\bibitem[\protect\citeauthoryear{Fan and Zhang}{Fan and
  Zhang}{2004}]{fan2004generalised}
Fan, J. and W.~Zhang (2004).
\newblock Generalised likelihood ratio tests for spectral density.
\newblock {\em Biometrika\/}~{\em 91\/}(1), 195--209.

\bibitem[\protect\citeauthoryear{Gaffney and Smyth}{Gaffney and
  Smyth}{2004}]{gaffney2004joint}
Gaffney, S.~J. and P.~Smyth (2004).
\newblock Joint probabilistic curve clustering and alignment.
\newblock In L.~Saul, Y.~Weiss, and L.~Bottou (Eds.), {\em Advances in neural
  information processing systems 17}, pp.\  473--480. Cambridge, MA: MIT Press.

\bibitem[\protect\citeauthoryear{Genolini and Falissard}{Genolini and
  Falissard}{2010}]{genolini2010kml}
Genolini, C. and B.~Falissard (2010).
\newblock Kml: k-means for longitudinal data.
\newblock {\em Computational Statistics\/}~{\em 25\/}(2), 317--328.

\bibitem[\protect\citeauthoryear{Green and Silverman}{Green and
  Silverman}{1993}]{green1993nonparametric}
Green, P.~J. and B.~W. Silverman (1993).
\newblock {\em Nonparametric regression and generalized linear models: a
  roughness penalty approach}.
\newblock CRC Press.

\bibitem[\protect\citeauthoryear{Hutchinson and De~Hoog}{Hutchinson and
  De~Hoog}{1985}]{hutchinson1985smoothing}
Hutchinson, M.~F. and F.~De~Hoog (1985).
\newblock Smoothing noisy data with spline functions.
\newblock {\em Numerische Mathematik\/}~{\em 47\/}(1), 99--106.

\bibitem[\protect\citeauthoryear{Jacques and Preda}{Jacques and
  Preda}{2013}]{jacques2013funclust}
Jacques, J. and C.~Preda (2013).
\newblock Funclust: A curves clustering method using functional random
  variables density approximation.
\newblock {\em Neurocomputing\/}~{\em 112}, 164--171.

\bibitem[\protect\citeauthoryear{James, Hastie, and Sugar}{James
  et~al.}{2000}]{james2000principal}
James, G.~M., T.~J. Hastie, and C.~A. Sugar (2000).
\newblock Principal component models for sparse functional data.
\newblock {\em Biometrika\/}~{\em 87\/}(3), 587--602.

\bibitem[\protect\citeauthoryear{James and Sugar}{James and
  Sugar}{2003}]{james2003clustering}
James, G.~M. and C.~A. Sugar (2003).
\newblock Clustering for sparsely sampled functional data.
\newblock {\em Journal of the American Statistical Association\/}~{\em
  98\/}(462), 397--408.

\bibitem[\protect\citeauthoryear{Jones and Nagin}{Jones and
  Nagin}{2007}]{jones2007advances}
Jones, B.~L. and D.~S. Nagin (2007).
\newblock Advances in group-based trajectory modeling and an sas procedure for
  estimating them.
\newblock {\em Sociological Methods and Research\/}~{\em 35\/}(4), 542--571.

\bibitem[\protect\citeauthoryear{Krivobokova and Kauermann}{Krivobokova and
  Kauermann}{2007}]{krivobokova2007note}
Krivobokova, T. and G.~Kauermann (2007).
\newblock A note on penalized spline smoothing with correlated errors.
\newblock {\em Journal of the American Statistical Association\/}~{\em
  102\/}(480), 1328--1337.

\bibitem[\protect\citeauthoryear{Lin, Hennig, and Huang}{Lin
  et~al.}{2015}]{lin2015clustering}
Lin, C., C.~Hennig, and C.-L. Huang (2015).
\newblock Clustering and a dissimilarity measure for methadone dosage time
  series.
\newblock In {\em Proceedings of ECDA-2014, Bremen, Germany}, pp.\  to appear.
  Springer, Berlin.

\bibitem[\protect\citeauthoryear{Liu and Yang}{Liu and
  Yang}{2009}]{liu2009simultaneous}
Liu, X. and M.~C. Yang (2009).
\newblock Simultaneous curve registration and clustering for functional data.
\newblock {\em Computational Statistics and Data Analysis\/}~{\em 53\/}(4),
  1361--1376.

\bibitem[\protect\citeauthoryear{Maharaj}{Maharaj}{1996}]{maharaj1996significance}
Maharaj, E.~A. (1996).
\newblock A significance test for classifying arma models.
\newblock {\em Journal of Statistical Computation and Simulation\/}~{\em
  54\/}(4), 305--331.

\bibitem[\protect\citeauthoryear{McNicholas and Murphy}{McNicholas and
  Murphy}{2010}]{mcnicholas2010model}
McNicholas, P.~D. and T.~B. Murphy (2010).
\newblock Model-based clustering of longitudinal data.
\newblock {\em Canadian Journal of Statistics\/}~{\em 38\/}(1), 153--168.

\bibitem[\protect\citeauthoryear{Montero and Vilar}{Montero and
  Vilar}{2014}]{montero2014tsclust}
Montero, P. and J.~A. Vilar (2014).
\newblock {TSclust}: An {R} package for time series clustering.
\newblock {\em Journal of Statistical Software\/}~{\em 62\/}(1), 1--43.

\bibitem[\protect\citeauthoryear{Murtagh and Contreras}{Murtagh and
  Contreras}{2012}]{murtagh2012algorithms}
Murtagh, F. and P.~Contreras (2012).
\newblock Algorithms for hierarchical clustering: an overview.
\newblock {\em Wiley Interdisciplinary Reviews: Data Mining and Knowledge
  Discovery\/}~{\em 2\/}(1), 86--97.

\bibitem[\protect\citeauthoryear{{R Core Team}}{{R Core Team}}{2016}]{CRAN}
{R Core Team} (2016).
\newblock {\em R: A Language and Environment for Statistical Computing}.
\newblock Vienna, Austria: R Foundation for Statistical Computing.

\bibitem[\protect\citeauthoryear{Ramaswamy, Rastogi, and Shim}{Ramaswamy
  et~al.}{2000}]{ramaswamy2000efficient}
Ramaswamy, S., R.~Rastogi, and K.~Shim (2000).
\newblock Efficient algorithms for mining outliers from large data sets.
\newblock In {\em ACM SIGMOD Record}, Volume~29, pp.\  427--438. ACM.

\bibitem[\protect\citeauthoryear{Ramsay and Silverman}{Ramsay and
  Silverman}{2005}]{ramsay2005smoothing}
Ramsay, J. and B.~Silverman (2005).
\newblock Smoothing functional data with a roughness penalty.
\newblock In {\em Functional Data Analysis}, pp.\  81--109. Springer.

\bibitem[\protect\citeauthoryear{Tarpey and Kinateder}{Tarpey and
  Kinateder}{2003}]{tarpey2003clustering}
Tarpey, T. and K.~K. Kinateder (2003).
\newblock Clustering functional data.
\newblock {\em Journal of classification\/}~{\em 20\/}(1), 093--114.

\bibitem[\protect\citeauthoryear{Vilar, Alonso, and Vilar}{Vilar
  et~al.}{2010}]{vilar2010non}
Vilar, J.~A., A.~M. Alonso, and J.~M. Vilar (2010).
\newblock Non-linear time series clustering based on non-parametric forecast
  densities.
\newblock {\em Computational Statistics and Data Analysis\/}~{\em 54\/}(11),
  2850--2865.

\bibitem[\protect\citeauthoryear{Wahba and Wendelberger}{Wahba and
  Wendelberger}{1980}]{wahba1980some[TPS]}
Wahba, G. and J.~Wendelberger (1980).
\newblock Some new mathematical methods for variational objective analysis
  using splines and cross validation.
\newblock {\em Monthly weather review\/}~{\em 108}, 1122--1143.

\bibitem[\protect\citeauthoryear{Wang}{Wang}{1998}]{wang1998smoothing}
Wang, Y. (1998).
\newblock Smoothing spline models with correlated random errors.
\newblock {\em Journal of the American Statistical Association\/}~{\em
  93\/}(441), 341--348.

\bibitem[\protect\citeauthoryear{Warren~Liao}{Warren~Liao}{2005}]{warren2005clustering}
Warren~Liao, T. (2005).
\newblock Clustering of time series data--a survey.
\newblock {\em Pattern recognition\/}~{\em 38\/}(11), 1857--1874.

\bibitem[\protect\citeauthoryear{Yassouridis}{Yassouridis}{2016}]{funcy}
Yassouridis, C. (2016).
\newblock {\em funcy: Functional Clustering Algorithms}.
\newblock R package version 0.8.4.

\end{thebibliography}

\end{document}